\documentstyle[psfig]{mn}

\footnotesize
\newdimen\digitwidth    
\setbox0=\hbox{\rm0}
\digitwidth=\wd0
\catcode`!=\active
\def!{\kern\digitwidth}
\normalsize

\title[Finding binary and millisecond pulsars] {The Parkes
Multibeam Pulsar Survey: V. Finding binary and millisecond
pulsars} \author[A.~J. Faulkner et al.]  {
A.~J.Faulkner$^1$,\thanks{Email: afaulkne@jb.man.ac.uk}
I.~H. Stairs$^2$, M. Kramer$^1$, A.~G. Lyne$^1$, G. Hobbs$^3$,
\newauthor A. Possenti$^4$, D.~R. Lorimer$^1$, R.~N. Manchester$^3$,
M.~A. McLaughlin$^1$, N. D'Amico$^{4,5}$, \newauthor F. Camilo$^6$,
M. Burgay$^4$ \\ $^1$ University of Manchester, Jodrell Bank
Observatory, Macclesfield, Cheshire, SK11~9DL, UK\\ $^2$ Department of
Physics \& Astronomy, University of British Columbia, 6224
Agricultural Road, Vancouver, B.C. V6T 1Z1, Canada \\ $^3$ Australia
Telescope National Facility, CSIRO, P.O.~Box~76, Epping NSW~1710,
Australia\\ $^4$ INAF - Osservatorio Astronomico di Cagliari,
Loc. Poggio dei Pini, Strada 54, 09012, Capoterra (CA), Italy \\ $^5$
Universita' degli Studi di Cagliari, Dipartimento di Fisica, SP
Monserrato-Sestu km 0,7, 90042, Monserrato (CA), Italy \\ $^6$
Columbia Astrophysics Laboratory, Columbia University, 550 West 120th
Street, New York, NY~10027, USA\\ }

\date{2004 Aug 11}
\begin{document}

\maketitle
\newcommand{\setthebls}{
}

\setthebls

\begin{abstract}
The Parkes Multibeam Pulsar Survey is the most successful survey of
the Galactic plane ever performed, finding over 600 pulsars in the
initial processing. We report on reprocessing of all 40,000 beams with
a number of algorithms, including conventional frequency-domain
searches and an acceleration search for fast binary pulsars. The very
large volume of results coupled with the need to distinguish new
candidates from known pulsars and their many harmonics, often with
multiple detections from different search algorithms, necessitated the
development of a new graphical selection tool tightly linked to a
web-based results database. We discuss and demonstrate the benefits of
these software systems which are specifically designed for large
survey projects. The results of this processing have been encouraging;
we have discovered 128 new pulsars including 11 binary and 15
millisecond pulsars, in addition to those previously found in the
survey, we have thus far discovered 737 pulsars. In this paper we
discuss the discoveries of PSR J1744$-$3922, a 172~ms mildly recycled
pulsar in a 4.6~hr orbit that exhibits nulling behaviour, not
previously observed in recycled or binary objects; PSR J1802$-$2124,
an intermediate mass binary pulsar (IMBP) and PSR J1801$-$1417, a
solitary millisecond pulsar.

\end{abstract}

\begin{keywords}
pulsars: general --- pulsars: searches --- pulsars: timing
--- pulsars: individual PSR J1744$-$3922
--- pulsars: individual PSR J1801$-$1417
--- pulsars: individual PSR J1802$-$2124
\end{keywords}

\section{Introduction}

Binary pulsars provide the only environment for precise measurements
to test general relativity in strong-field conditions (Taylor
1993\nocite{tay93a}). The orbital decay of double neutron star systems
is currently the only observational demonstration of the existence of
gravitational waves (Taylor \& Weisberg 1989\nocite{tw89}). For the
effects of general relativity to be seen, the pulsar usually has to
have a massive companion ($>$ 1~M$_{\odot}$) and an orbital period of
only a few hours. The discovery of a black hole -- pulsar binary would
be very important for even more stringent tests of general relativity
(Damour \& Esposito-Far\`{e}se 1998\nocite{de98}). The pulsar itself
must be timed to high precision, implying that ideally it will be a
strong millisecond pulsar with a narrow pulse profile.

Millisecond pulsars with spin periods less than 2~ms probe the
equation of state for the degenerate matter which makes up neutron
stars. PSR B1937+21 has the shortest-known period of 1.55~ms (Backer
et al. 1982\nocite{bkh+82}). Finding a new pulsar with a shorter spin
period would clearly be very significant. Conversely, not finding such
a system in a major survey reinforces the so-called stiff theories of
equation of state (e.g. Friedman 1995\nocite{fri95}) or other
spin-period limitations, such as gravitational radiation (Andersson,
Kokkotas \& Schutz 1999\nocite{aks99}).

Finding binary pulsars with orbital periods of a few hours or less is
very difficult; the received signal from pulsars in binary systems has
a period variation dependent upon the orbital motion, making
conventional spectral analysis considerably less sensitive. To recover
sensitivity to binary pulsars, particularly in relatively long
observations, requires the use of computationally expensive
`acceleration' searches (e.g. \nocite{clf+00} Camilo et al. 2000a).

We report here the successful use of an efficient acceleration search
algorithm and optimised interference filters applied to the large
number of observations taken as part of the Parkes Multibeam Pulsar
Survey (PMPS), described in Manchester et al. (2001)\nocite{mlc+01},
hereafter `Paper I'.  Section \ref{sec:pmps} briefly summarizes the
PMPS, with a review of all employed search algorithms in Section
\ref{sec:algorithms}. Section \ref{sec:processing} details the data
reduction processing and Section \ref{sec:reaper} describes a
graphical approach to reviewing the results and ensuring that known
pulsars were correctly identified. In Section \ref{sec:performance} we
demonstrate that the system performed as expected.  Also, we report
the discovery of a significant number of millisecond pulsars using
conventional processing techniques.  These discoveries are discussed
in Section \ref{sec:discoveries} and future activities are considered
in Section \ref{sec:discussion}.

\section{Parkes Multibeam Pulsar Survey system}
\label{sec:pmps}

The PMPS covers the Galactic plane in the region $|b|<5^o$ and $260^o
<l< 50^o$. It has been very successful, with reported discoveries of
more than 600 new pulsars so far, discussed in Paper I, Morris et
al. (2002)\nocite{mhl+02}, Kramer et al. (2003)\nocite{kbm+03} and
Hobbs et al. (2004)\nocite{hfs+04}. However, there were only four
millisecond (which we define as having a period $<$ 30~ms) and twelve
binary pulsars reported. In 2002 a new Beowulf computer cluster at
Jodrell Bank Observatory, COBRA, started to be used to process all the
data from the PMPS, including some data that had never been processed
before, employing newly developed algorithms (hereafter `COBRA
processing').

The details of the receiving system and its performance can be found
in Paper I. It is the most sensitive survey of the Galactic plane ever
undertaken, consisting of 35-min integrations performed with a
sensitive, wide bandwidth receiver. Most of the observations took
place between 1997 and 2002. The data now represent a unique archive,
which may not easily be reproduced due to steadily increasing radio
frequency interference. By processing the data in a consistent
fashion, all pulsars above the limiting flux density thresholds within
the survey area can readily be compared. The raw (as recorded at the
telescope) data, the processing results and plots are archived using
on-line disk storage.  These data are expected to provide a basis for
further analyses (e.g. detailed population studies of Galactic
pulsars).

There are $\sim$14\% more observations than the 2670 pointings of 13
beams originally planned. Additional observations were taken for a
number of reasons including aborted observations due to high winds,
difficulty in reading some tapes thus losing some data, or
administrative errors. The additional observations can be useful
duplicates of the main observations; for example they can be used for
initial validation of a candidate or a discovery of a pulsar that has an
intermittent received signal, possibly because of nulling (see
\nocite{bac70} Backer et al., 1970) or scintillation (see
\nocite{lr68} Lyne \& Rickett, 1968). All these data, including truncated
observations (17.5 -- 35~mins), have now been processed.

\section{Search Algorithms}
\label{sec:algorithms}

The standard frequency domain search using a fast Fourier transform
(FFT) is extremely good at detecting stable periodicities from
solitary pulsars. The standard FFT approach was applied to all the data
of the PMPS as described in Paper I. Harmonic summing (Taylor \&
Huguenin 1969\nocite{th69}) was used to compensate for the relatively
low duty cycle of a pulsar to maximise sensitivity. However, there are
some classes of pulsars where the FFT-based search is less sensitive,
including: pulsars in binary systems, `long-period' pulsars which we
define as having periods $>$ 3~s and pulsars in which we can
observe only intermittent individual pulses. Throughout this paper we
use the term `detection' for any signals identified by a search
algorithm, including known pulsars or possible new pulsars and their
harmonics and interference. The term `candidate' is a detection that
has been reviewed and considered to be a possible new pulsar.

\subsection{Acceleration Searches}
\label{s:acceleration}

\begin{figure}
\vspace{0.0truecm}
\hrule
\vspace{0.3truecm}

\centerline{\psfig{file=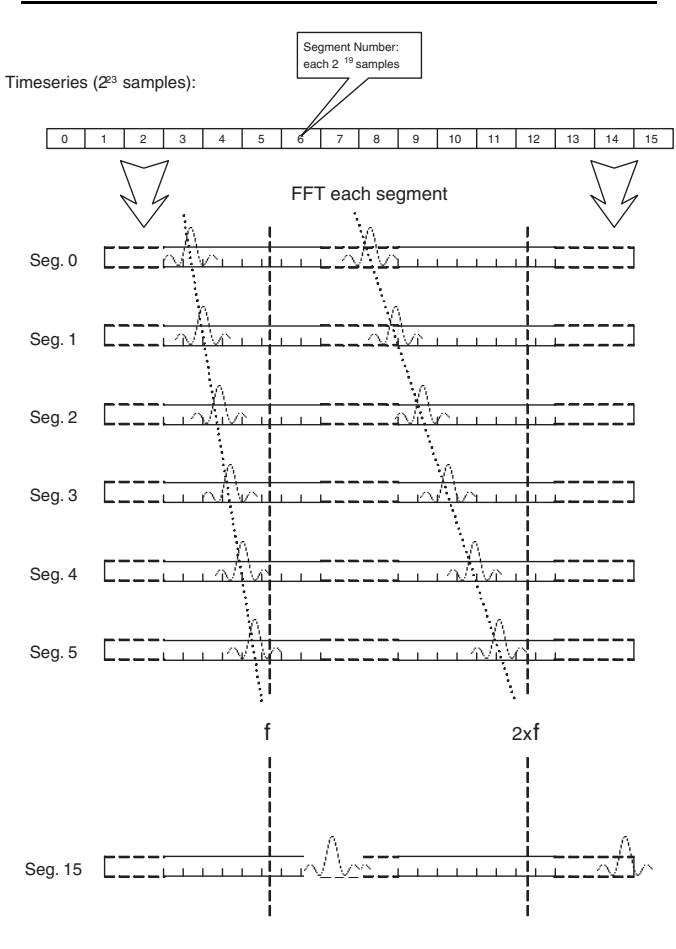,height=12truecm}}
\vspace{0.3truecm}
\hrule

\caption{\small An illustration of the stack search approach to
acceleration searching. The received time series is first broken into
a number of segments (in this case 16), which are individually FFT'd
and the amplitude spectrum formed, using interpolation to recover
signals lying midway between bins. These spectra are then summed at
different rates of change of frequency to produce a final
spectrum. Note that higher harmonics of signals need to have a rate of
change of frequency proportional to the harmonic number.}
\label{f:stack}
\end{figure}

Searching for pulsars in binary systems is made difficult because
their changing line-of-sight velocity causes a varying received pulsar
period. The effect on a search is to spread the signal detection over
a frequency range. An FFT divides the received signal into discrete
frequencies, so-called `spectral bins', hence, a varying received
frequency will spread a detection over a number of bins, thus reducing
significantly the signal-to-noise ratio (S/N) of a detection. This
effect becomes more severe with shorter orbital and pulsar periods,
and with longer observations. Consequently, some of the most
interesting systems, viz, millisecond pulsars in tight binary orbits, are
very difficult to detect.

It is possible to modify the time series such that the pulse period
variations caused by binary motion are removed; an FFT-based search
can then be used to find periodicities with no loss of
sensitivity. This is a `coherent' search, meaning that all the
available information (time, amplitude and phase) from an observation
is preserved and operated on as a whole. However, this method is
extremely computationally expensive in a blind search, with three
parameters (binary period, orbital amplitude and phase) to search for
circular orbits, with an additional two (orbital eccentricity and
longitude of periastron) for significantly elliptical orbits. This is
impractical using today's technology (e.g. Dhurander \& Vecchio
2001\nocite{dv01}). 

The task is simplified by only searching over one parameter, average
acceleration, which approximates the period changes over time to a
straight line, referred to as a `linear acceleration' search. This
works well for orbital periods which are much longer than the
observation length, particularly if the observations happen to be made
at an orbital phase where the pulsar's acceleration varies least from
being constant (Johnston \& Kulkarni 1991\nocite{jk91}). Linear
acceleration searches can be performed either in the time or frequency
domain. In the time domain, the time series is adjusted by resampling
to compensate for the quadratic change in pulse phase over the
observation time, and then an FFT based search is performed. This
needs to be repeated for all the different accelerations tested as
discussed in \nocite{clf+00} Camilo et al. (2000a).

Alternatively, in the frequency domain, an FFT of the zero
acceleration time series is performed and then a series of matched
filters is applied to correct the frequency spectrum to look for
accelerated periodicities (e.g. Ransom et
al. 2002\nocite{rem02}). While the frequency domain approach is
computationally more efficient for a blind search, both techniques
still require considerable processing at each dispersion measure (DM)
to be searched. Consequently, a coherent linear acceleration search is
generally only feasible if the DM is already known, as in the case of
globular cluster searches when there is already a confirmed pulsar. It
is currently impractical to apply a coherent acceleration search using
200 trials (see Camilo et al. 2000a\nocite{clm+00}) to the PMPS data,
since the total processing time would increase by $\sim$100 times.

By operating on the data in sections, known as an `incoherent' search,
some information is lost, in this case phase. However, by sacrificing
some sensitivity, an incoherent acceleration search, or `stack search'
as discussed by Wood et al. (1991\nocite{wnh+91}) can be used to save
substantial computing resources. The stack search is illustrated in
Figure \ref{f:stack}. It involves splitting the received time series
into a number of segments (16 in the case of the PMPS), FFT'ing each
of these individually and producing an amplitude spectrum. Use of very
short segment observations, over time $T_{\rm seg}$, considerably
improves the relative sensitivity to varying periods. In a short
observation the period does not change much and each spectral bin
covers a much larger frequency range ($\Delta f=1/T_{\rm seg}$). By
summing each of the bins of the same frequency from all the segments,
a spectrum of the full observation can be made. Some overall
sensitivity will be lost due to the incoherent summing. However, the
pulse duty cycle, hence the summing of harmonics makes it hard to
predict the precise effect on spectral S/N. In the case of the PMPS,
there are 5 trials, the fundamental only and then progressively adding
up to the 2nd, 4th, 8th and 16th harmonics. By comparing the spectral
S/N for the standard search with spectral S/N for the stack search for
a variety of solitary pulsars we found there to be a S/N loss of
between 15-35\% over a coherent search. Linear accelerations, $a$, can
then be applied to the whole sequence by adding up the bins at a slope
proportional to the change in frequency over the observation. The
frequency change, $\Delta \nu$, from segment to segment for an
observation of length $T$ with number of segments $N_{\rm seg}$ and
speed of light $c$ is:
\begin{equation}
\Delta \nu=\nu_0 aT/(N_{\rm seg}c) \rm{~Hz}.
\end{equation}
Since the number of bins shifted must be an integer, the frequency
change needs to be rounded to the nearest bin. Prior to harmonic
summing, the $n$th harmonic must be corrected for acceleration using
a slope $n$ times the fundamental slope. The FFTs for all of the
segments take a similar or less time to produce than a single full
length FFT. Spectra for each of the trial accelerations are produced
by addition, making the stack search fast and efficient. This is the
technique that has been applied to processing of the PMPS. Without the
stack search, the double neutron star PSR J1756-22\footnote{Names
quoted with just two digits of declination are interim.}
(\nocite{}Faulkner et al. in prep.) would not have been detected.

Highly accelerated systems which have more than 1.5 orbits over the
length of the observation can be found using a `phase modulation'
search (Jouteux et al. 2002\nocite{jrs+02}, Ransom, Cordes \&
Eikenberry 2003\nocite{rce03}). In the case of the PMPS this would
apply to orbital periods less than 20~min. The algorithm finds the
orbital periodicity by identifying sidebands in the power spectrum of
the time-series. Any sidebands present are summed by stepping through
the power spectrum taking short FFTs. This is a fast process and
should reveal the pulsar period and orbital period. It has been
incorporated into the search algorithms and any results will be
reported in a separate paper.

\subsection{Other search algorithms incorporated}

Frequency domain searches using FFT's are effective for most of the
range of pulse periods, from the Nyquist limit up to approximately 3 s
in the case of the PMPS. However, when dealing with periods which are
greater than 3~s, which lie in the first few thousand spectral bins,
the period resolution becomes very coarse. In addition, the presence
of low-frequency noise in the Fourier spectrum of the
radio-astronomical time series can be very difficult to remove. By
using a time domain search the period resolution is much finer and
resolves the true spin period more accurately. A time domain search
works effectively on arbitrary pulse profiles, therefore is more
effective on the very narrow pulses typical of long period
pulsars. Since the search is for long periods, greater than 3~s, the
sampling time can be considerably increased with no significant loss
of sensitivity, thus reducing the processing required. An efficient
algorithm (\nocite{sta69}Staelin 1969) has been included in the COBRA
processing; details of this implementation will be described in a
separate paper. This algorithm has been successful at finding a new
pulsar with a 7.7 s period, which will be published in a future paper.

It has been shown (\nocite{mc03}McLaughlin \& Cordes 2003) that some
pulsars can only be detected by observing their individual giant
pulses. It is possible to detect these individual pulses if they are
large enough. A `single pulse search' for dispersed irregular
pulses has been incorporated into the search routines. Over 30\% of
all pulsars (new and known) detected in the PMPS were also detected in
the single-pulse search. In addition, the single-pulse search has
detected four pulsars which were not detectable in the standard
periodic processing. The results from this search will be presented in
a future paper.

\section{Processing}
\label{sec:processing}

Details of the observing and search analysis systems are mostly as
reported in Paper I. Modifications and enhancements are discussed
below.

All the data were gathered at Parkes and stored on 160 DLT tapes. The
tapes have been collected at Jodrell Bank Observatory for processing.
With the ongoing reduction in disk storage costs, all the raw data
have been transferred to 5TB of on-line RAID disk systems. Thus,
alternative processing approaches can easily be tried in the future.

The data flow from observation data through processing to produce
detection lists and plot files is shown in Figure
\ref{f:cobra_proc}. The processing task is divided into two principal
areas: separating the data into individual beams and searching for
major interference signals, followed by detailed searches of
individual beams.

The first parts of the processing, stages one and two described in
Paper I, take the raw data files, split them into thirteen individual
beams, and perform interference search and excision. The resulting
datafiles, header files and accounting files are passed to the search
stage.

During the period of PMPS observations and on-going processing,
 progressively more frequency filters were added to combat
 interference. However, they also restricted the amount of spectrum
 left for searching, particularly at millisecond pulsar
 frequencies. Significant rationalisation of these `interference'
 frequency bands was made prior to this COBRA processing, which
 substantially increased the amount of spectrum available to find new
 pulsars. There is now $<$~10\% of the spectrum removed, as opposed to
 $\sim$40\% in previous analyses (Hobbs 2002)\nocite{hob02}. This is a
 significant contributor to the success at finding a high proportion
 of new millisecond pulsars. A summary of the filters is shown in
 Table \ref{t:filters}; a full discussion of the purpose and changes
 made to each of the filters is given by Hobbs (2002)\nocite{hob02}.

\begin{table}
\begin{small}
\begin{center}
\caption{\small Frequency filters for removing identified
interference, probably related to mains 50~Hz plus harmonics and local
and instrumental sources.  $\nu$ is the central frequency of the
individual filter, $i$ and $j$ are integers and $\nu_{\rm nyq}$ is the
Nyquist frequency of the PMPS.  In all cases $\nu$ wraps around
the Nyquist frequency and returns to lower frequencies.}

\begin{tabular}{llll}

\hline  
\hline
Central frequency $\nu$ & Width & Ranges  & Limit \\
   (Hz)           & (Hz) & ($i,j$)  & (Hz) \\ 
\hline
$\frac{2^{(i-1)}}{8.192}\times(1+2^{j-1})$ 
      & 0.001 & $ \{^{i:1\rightarrow15}_{j:1\rightarrow10}$ 
      & $\nu < \nu_{\rm nyq}$ \\ 

$\nu_2 \times i$ &0.1& $0\rightarrow120$& $\nu > 2$ \\

$\left|\nu _1 \times i + \nu _2 \times j \right|$ & 0.004 
      & $\{^{i:1\rightarrow50}_{j:-20\rightarrow20}$ & $\nu > 1.00$ \\

$\nu_1 \times i$ & 0.04 & $51\rightarrow160$ &  \\

$\nu_1 \times i$ & 0.4 & $1\rightarrow \nu_{\rm nyq4}/\nu_1$&  \\

$\nu_1 \times i + 3.9$ & 0.3 & $1\rightarrow \nu_{\rm nyq2}/\nu_1$ & \\
$\nu_1 \times i - 3.9$ & 0.3 & $1\rightarrow \nu_{\rm nyq2}/\nu_1$ & \\

$50 \times i$ & 0.2 & $1\rightarrow \nu_{\rm nyq}$ &  \\
$200+i$ & 0.26 & $-20\rightarrow20$ &  \\
$0.14704 \times i$ & 0.0024 & $1\rightarrow100$ &  \\

$1349$            & 12 \\
$508.65 \times i$ & $0.400 \times i$ & $1 \rightarrow 2$ & \\
$2.045$           & 0.002 \\
$1.2032 \times i$ & $0.001 \times i$ & $1 \rightarrow 40$ & \\         
$1.0000 \times i$ & $0.001 \times i$ & $1 \rightarrow 30$ & \\         
$0.6433 \times i$ & $0.001 \times i$ & $1 \rightarrow 50$ & \\         
     
\hline
\label{t:filters}
\end{tabular}
\end{center}
\begin{tabular}{lll}
Frequency definitions:& $\nu_1 = 134.99776$\,Hz, & $\nu_2 = 99.99592$\,Hz, \\
$\nu_{\rm nyq}=2000$\,Hz,  & $\nu_{\rm nyq2}=4000$\,Hz,   
& $\nu_{\rm nyq4}=8000$\,Hz \\
\end{tabular}
\end{small}
\end{table}

Terrestrial interference is normally not dispersed, so an FFT is
taken of the time series not adjusted for dispersion, the `zero-DM'
spectrum, and the result is stored. These data can be used for a
future analysis of interference signals at Parkes.

\subsection{Beam processing}
\label{sec:beam_proc}

The searching of each of the 40,000 beams (stage 3 of Paper I) can be
done on a beam-by-beam basis, essentially unaffected by any other data
or process. This is the perfect environment for a Beowulf cluster,
which in its simplest form is a collection of interconnected
PC's. Hence, each beam is assigned to a separate processor; each node
consisting of two processors is able to search two beams
simultaneously. An individual beam was typically processed in about
6$-$7 hours on a single processor.  Excellent total throughput of
processing is achieved by assigning many processors, typically about
100. A tape of $\sim$300 beams was processed in a day.

\begin{figure}
\centerline{\psfig{file=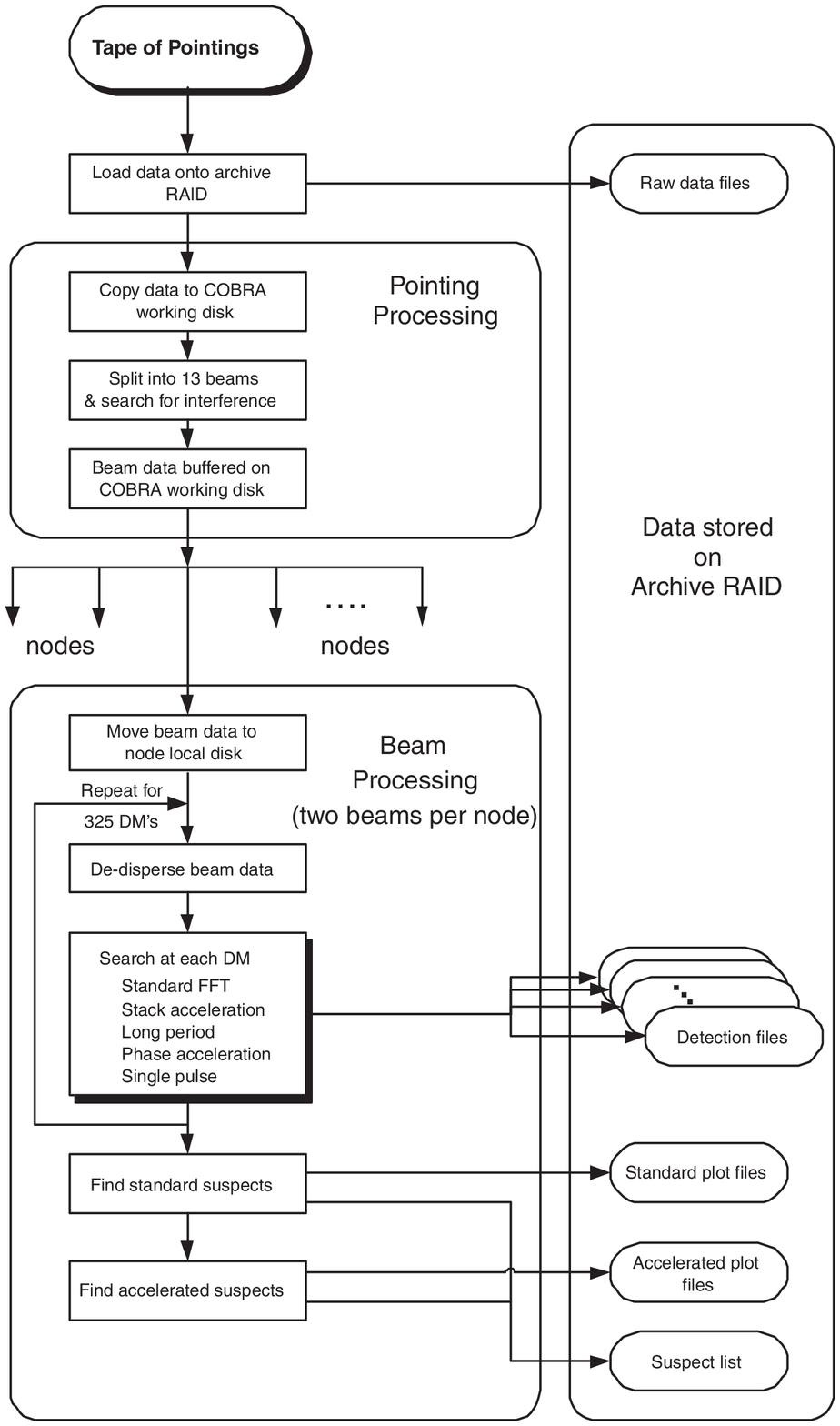,height=14truecm,angle=0}}
\vspace{0.0truecm}
\caption{\small Processing on COBRA. There are 3 phases: firstly, data
from the tape are loaded onto archive RAID; the data are then copied to
COBRA's local working disk and split into beams followed by a
search for interference specific to a tape; the beam data, headers and
accounting files are put into a buffer; the beam processing searches
an individual beam on one processor, followed by candidate
identification.}
\label{f:cobra_proc}
\end{figure}

The various tasks (see Figure \ref{f:cobra_proc}) that make up a
search of a beam are controlled by a script running on each
processor.

\begin{figure*}
\hrule
\vspace{0.3truecm}
\centerline{\psfig{file=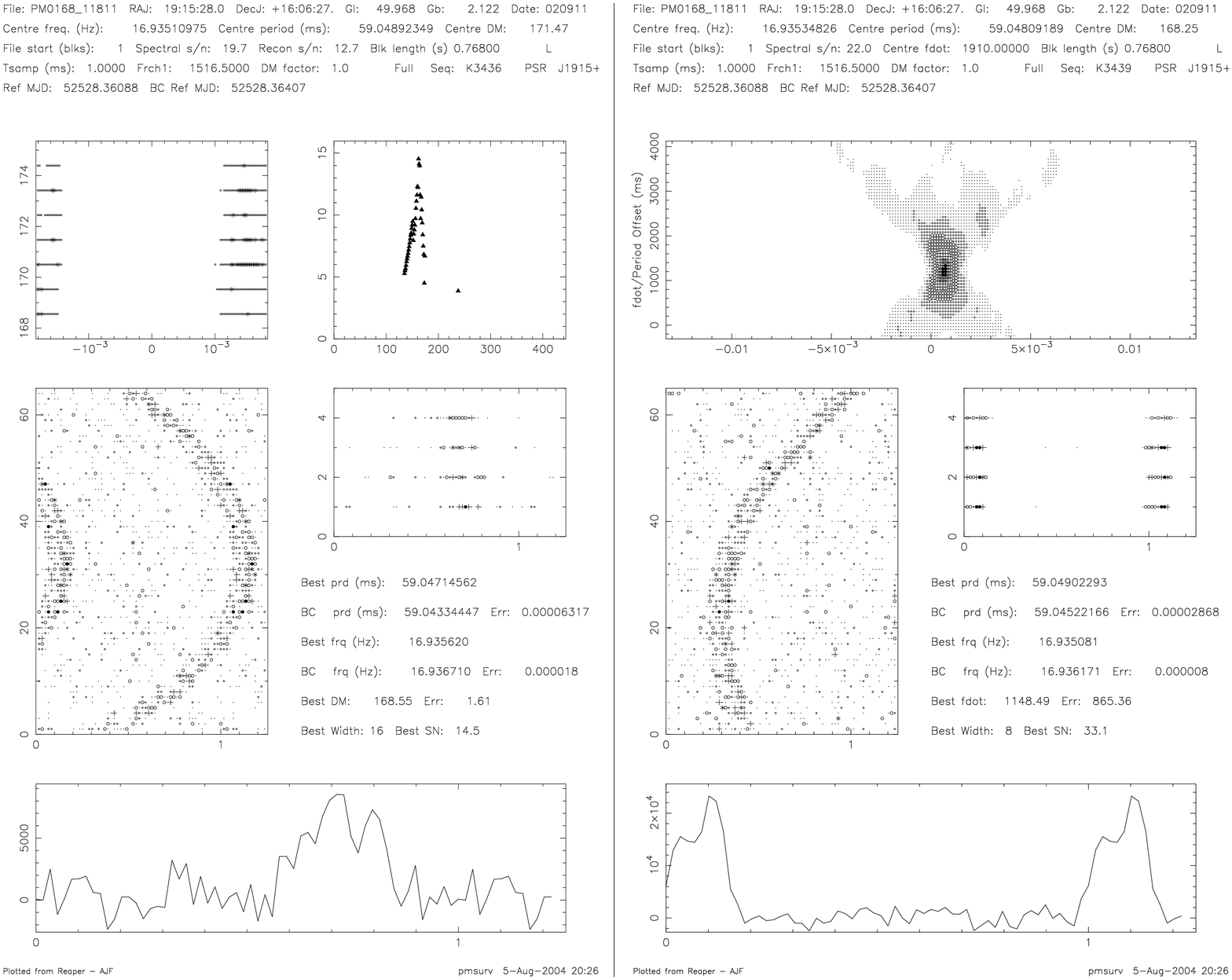,height=14truecm}}
\vspace{0.3truecm}
\hrule

\caption{\small Example candidate plots comparing standard (left) and
acceleration (right) searches, these show the re-detection of PSR
B1913+16, the first binary pulsar. It can be seen that the
acceleration search considerably improves the final S/N and pulse
profile.The acceleration information presented shows beam and
detection information at the top; the top plot shows the S/N variation
with small changes in the frequency due to acceleration (fdot) and
period; centre left is a signal strength plot over pulse period
(x-axis) and time through the observation (y-axis) of 64
sub-integrations; centre right shows four frequency bands plotted
against pulse phase; finally the corrected pulse profile is along the
bottom. The standard search information is similar to the accelerated
plot except the top left plot shows S/N variation with period changes
and DM, the top right plot shows S/N variation with widely changing
DM.}
\label{f:pdmplot}
\end{figure*}

The data are de-dispersed as described in Paper I. The search routines
then process the time-series for each of the 325 different DMs. The
spectra from each DM for both accelerated and standard FFT searches
are searched for the top 100 detections (increased from 50 in Paper
I), which ensures that all detections above the noise floor are
stored. The resulting detection lists are stored. To qualify the
detections, they are analysed for the highest S/N signals
(reconstructed by inverse transformation to the time domain for
standard searches and spectral S/N for the segmented search) and the
DM of the strongest detection is found for each identified
period. Detections with harmonic relationships are identified and
eliminated except the period corresponding to the strongest signal and
the detection with twice this period. Any detections over a S/N of 6
are then further analysed in the time domain largely as described in
Paper I. The standard FFT detections are searched over DM and a narrow
period range, using 64 sub-integrations (increased from 16
sub-integrations in Paper I) each of $\sim$30~s, to find an optimised
S/N and the results recorded, known as a `candidate plot'. The S/N is
calculated by convolving the profile with a box-car function, trialed
with widths from 1 bin to half the length of the profile. Similarly,
the best detections from the accelerated search are searched in period
and frequency derivative by repeated adjustment of the time series,
and the results also recorded. Examples of both types of candidate
plot are shown in Figure \ref{f:pdmplot}.

For the long-period search, six results from each beam which
are the best S/N detections from any DM searched, are recorded. These
are post-processed into candidate plots and stored for later
inspection. The single pulse search stores S/N, time index in the
observation, width and DM for all single pulses with a S/N in excess
of 4. The period, orbital period and DM information from the phase
modulation search are recorded for off-line analysis.

\vspace{0.5truecm}

\section{Finding Candidates}
\label{sec:reaper}

The processing produced a large quantity of output to be analysed,
with each tape of twenty six pointings producing around 50,000
qualified detections. These detections are a mixture of interference
signals, known pulsars with their integer, non-integer harmonics and
sub-harmonics and, occasionally, candidates but also their harmonically
related detections. Many of these results will be duplicated in more
than one of the standard, stack or long period search algorithms.

\subsection{Benefits of a graphical interface}

\begin{figure}
\centerline{\psfig{file=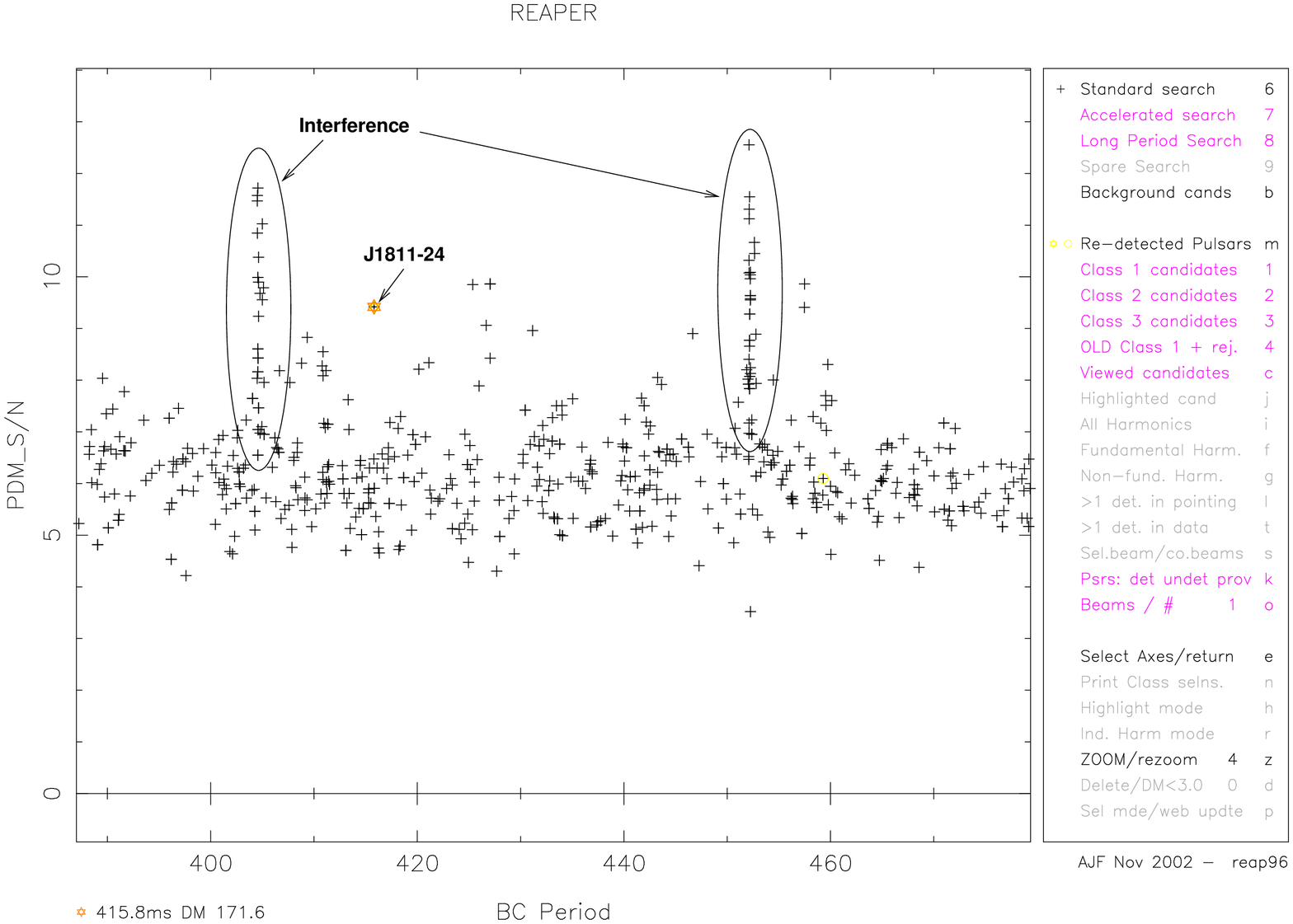,height=6truecm,angle=0}}
\vspace{0.0truecm}
\caption{\small Annotated \sc Reaper \normalfont screen-shot of the
discovery of PSR J1811-24 showing that the S/N is only slightly
above the noise floor and surrounded by interference, but still easily
discernable. Other isolated candidates were seen to be interference
from their candidate plots. }
\label{fig:1811_plot}
\end{figure}

With tens of thousands of signals to choose from, it is clearly not
feasible to manually inspect all the output plots from the search
algorithms. Two different philosophies to select a more reasonable
number of candidates for inspection are: (a) an algorithm that selects
mainly on S/N; (b) give the observer enough information in a
comprehensible form to enable a selection to be made. The algorithmic
approach was used originally on the PMPS as discussed in Paper
I. Using a graphical plot to visually select suitable candidates means
that interference is more easily avoided and fewer plots need to be
inspected. Candidates that would not be identified by a simple
algorithm can be seen graphically. For example, if the detection does
not have a high S/N, but is clear of interference and the noise floor,
which can relatively easily be seen on a graphical display, then it
would be selected for inspection, as with the discovery of PSR
J1811-24 shown in Figure \ref{fig:1811_plot}. These candidates may
be the most interesting discoveries. Other benefits of a graphical
approach are the relative ease of investigations of possible
candidates/known pulsars, e.g., detections in adjacent beams, harmonic
relationships, positions of known pulsars relative to beam centres
etc. The graphical approach has been used successfully for the
Swinburne intermediate-latitude pulsar survey (see \nocite{ebsb01}
Edwards et al. 2001).

\subsection{\sc{Reaper}}

\begin{figure}
\centerline{\psfig{file=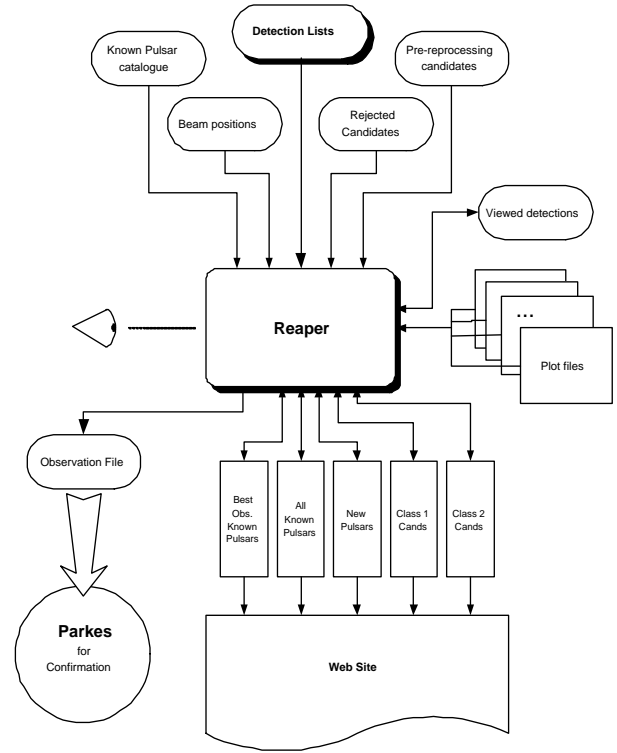,height=10truecm,angle=0}}
\vspace{0.0truecm}
\caption{\small Data flow after processing, through \sc Reaper
\normalfont, to create new candidates for observation at Parkes and
placing on a website. See text for `Best Obs.' description.}
\label{f:reap_flow}
\end{figure}

\begin{figure}
\centerline{\psfig{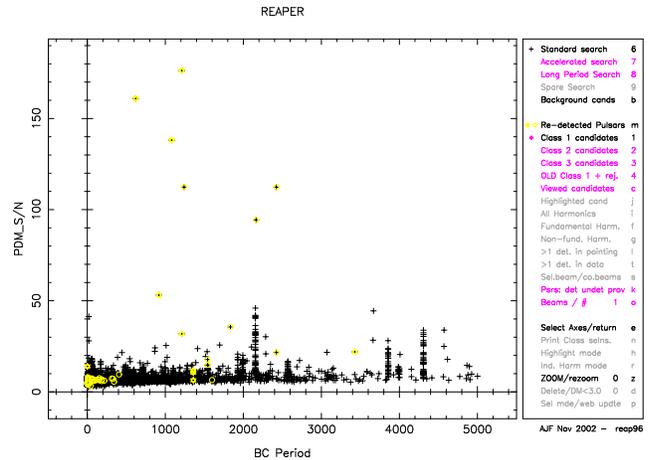}}
\vspace{0.0truecm}
\caption{\small A typical search screen for \sc Reaper \normalfont,
showing detections, candidates and known pulsars for a particular tape
of observations. The clickable menu of commands on the right highlights chosen
selections; the details are not relevant here. }
\label{f:reap_gen}
\end{figure}

A suite of programs, known as \sc Reaper\normalfont, for graphically
investigating the results of the PMPS has been developed. All the
results from one tape of observations are loaded at a time. The flow
of data from lists of detections and their candidate plots, existing
pulsars, previous candidates and previously viewed candidates through
to observation lists at Parkes and information on a web site is shown
in Figure \ref{f:reap_flow}. An illustration screen is shown in Figure
\ref{f:reap_gen}. Its user interface consists of an X-Y plot of a pair
of chosen parameters, typically period and S/N, with a menu of
selectable commands on the right. The parameters that may be plotted
are:

\begin{itemize}
\item spin period
\item spectral S/N
\item reconstructed S/N (if available)
\item candidate plot S/N
\item dispersion measure
\item pulse width
\item number of harmonics summed
\item spin frequency
\item spin frequency derivative (if available)
\item Galactic latitude and longitude
\end{itemize}

Other information can be superimposed with the detections, including
known pulsars and their harmonics, previously selected candidates and
previously viewed detections. This is done using a variety of colours
and symbols.

\begin{figure}
\centerline{\psfig{file=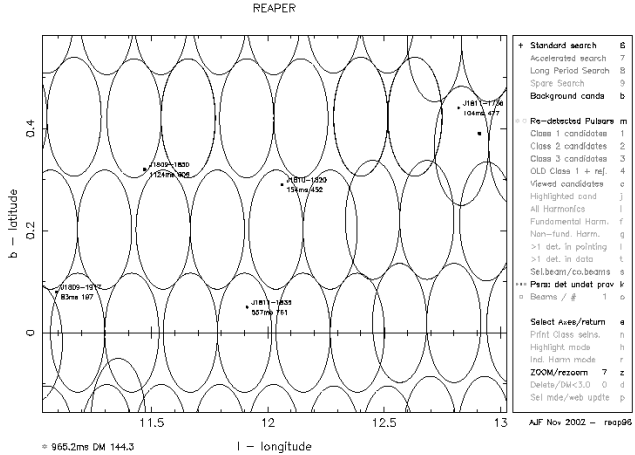,height=6truecm,angle=270}}
\vspace{0.0truecm}
\caption{\small \sc Reaper \normalfont displaying a Galactic plot
indicating FWHM of the observation beams and highlighting known
pulsars.}
\label{f:reap_gal}
\end{figure}

The X-Y plot can be navigated with zoom and movement commands. Any
detection can have its candidate plot viewed by selecting its
symbol. The Galactic position (see Figure \ref{f:reap_gal}) of a
possible candidate can be displayed. Also on the plot, the exact
position, period and DM of known pulsars can be shown, primarily to
confirm that there is no potential link between the detection and a
known object. Visual guides for the beam width and beam
identifications can be overlaid, hence showing the beams which may
have other detections.

If necessary, all the detections for a beam can be viewed, for
example, to investigate an apparent non-detection of a known pulsar or
see a search's response to a very large interference signal or
powerful nearby pulsar. All the detections from the standard and
accelerated searches are available. In this case the top 100
detections at each DM and harmonic sum are recorded with S/N and, if
appropriate, the acceleration in terms of frequency derivative. This
amounts to more than 300,000 detections per beam. Graphically these
can be viewed and zoomed into for particular periods over the various
DMs seen.

The graphical access to the results database for the entire PMPS is a
valuable tool for investigating known pulsars, interference signals
and potential single-pulse candidates.

\vspace{0.5truecm}

\vspace{1.0truecm}

\vspace{0.5truecm}

\subsection{Selecting and confirming candidates}

The selection of candidates is a matter of experience; in general
candidate plots are viewed if they have S/N $>$8, are not
identified as the fundamental or harmonic of a known pulsar and are
clear of detections in multiple beams, which would probably be
interference. Solitary candidates normally are present in both the
standard and acceleration searches. If the plot shows a reasonably
good profile at a DM which is greater than 3 pc cm$^{-3}$ (to avoid
terrestrial interference) and does not get changed substantially from
the spectral search through time-domain refinement (see Section
\ref{sec:beam_proc}) then it may be a candidate. It is important to
verify that it is not related to a known pulsar, so, a check is made
of the immediate area of sky $\pm1^o$ using \sc Reaper\normalfont's
Galactic plot. A final check is made by restricting a S/N -- period
plot to just the detection beam and ensuring that there are no large
interference signals in the beam which could cause spurious
detections.

After carefully considering a detection to ensure that it is unlikely
to be interference or a known pulsar, it may be assigned to one of
three classes as a candidate. Class 1 candidates will be prepared for
confirmation observations at Parkes at the next opportunity; class 2
candidates are kept for further consideration and class 3 candidates
are effectively discarded but kept for future reference. An
individual detection can be promoted or demoted by reselecting their
class within \sc Reaper.\normalfont

At any time, candidates selected are in various stages of the
confirmation process - waiting for confirmation, observations made but
not seen, gridded (e.g Morris et al.  2002\nocite{mhl+02}), confirmed
as a new pulsar or rejected. To keep all the information readily
available, \sc Reaper \normalfont writes out candidates to a web site;
other information is also collected from the central archive files and
displayed. The candidate plot is an essential feature, so, confirmation
plots are also displayed for ease of reviewing at the participating
institutions.

\sc Reaper \normalfont has been used to view all the processing
results from the PMPS, typically only taking about one hour per
tape. Both the processing system and \sc Reaper \normalfont are
modular and can be ported to other major surveys. For example, both
systems have been ported to the related survey, the Parkes High
Latitude Pulsars Survey (the PH Survey). It was in the PH Survey that
\sc Reaper \normalfont first detected the millisecond pulsar PSR
J0737$-$3039A of the double-pulsar system (Burgay 2004\nocite{bur04}).

\vspace{0.5truecm}

\section{System Performance}
\label{sec:performance}

With a complex system such as the PMPS, including the subsequent
processing, it is important to know that it performs as predicted. To
verify if the sensitivity, sky coverage and detection algorithms are
working correctly, we consider the re-detections of known pulsars,
pulsars in the search area which were not detected and any detections
that showed any anomalies.

There are more than 1000 pulsars in the survey area found from many
surveys including all the PMPS processing. As has been reported in
\nocite{hfs+04} Hobbs et al. (2004), of the 264 pulsars in the survey
area discovered by other surveys, all but eleven have been
re-detected with no unaccountable non-detections.


Pulsars were detected over the whole area of the survey at spin
periods from 1.85~ms to 7.7~s and at DM's from 3 to 1200 pc
cm$^{-3}$. Reviewing flux density measurements from Hobbs et
al. (2004) \nocite{hfs+04} of the weakest detected pulsars shows that
the system meets the sensitivity calculated in Paper I.

\subsection{Non-standard detections}

\begin{table}
\begin{small}
\begin{center}
\caption{\small Provisional names and parameters for all the binary
and millisecond pulsars found using COBRA processing, showing spin
period, orbital period and DM. The \it spectral \normalfont S/N for
the standard search (Std.) and the stack acceleration (Acc.) search
are shown to highlight the differences in search algorithm. The `Plot'
S/N is the refined S/N from the candidate plot, the values in \it
italics \normalfont are where the acceleration plot S/N has been
used.}

\begin{tabular}{lrrrrrr}
\hline
\hline
PSR J$^{\dag}$   & Period & Orbit & \multicolumn{1}{c}{DM} & 
\multicolumn{3}{c}{S/N} \\
	  	& (ms)          &
\multicolumn{3}{r}{(days)\hspace{0.5mm}(pc\,cm$^{-3}$)}\hspace{1.0mm}
Std. & Acc.& Plot \\ 
\hline
1125$-$60   &   2.630   &   8.75    &  53.1  & 11.5 & --   &  9.8 \\
1215$-$64   &   3.540   &   4.08    &  47.7  & 19.5 & 16.1 & 16.6 \\	
1439$-$54   &  28.639   &   2.12    &  14.5  &  7.1 & 10.1 & \it {17.5} \\
1551$-$49   &   6.284   &  --       & 114.6  & 12.2 & --   & 10.2 \\
1723$-$28   &   1.856   & tbd$^*$   &  19.9  & 16.7 & 40.3 & \it {20.8} \\ 
1726$-$29   &  27.081   &  --       &  60.9  &  8.9 &  7.3 & 12.2 \\
1744$-$3922 & 172.449   &   0.19    & 145.7  & 15.3 & 13.8 & \it {26.5} \\
1756$-$22   &  28.456   &   0.32    & 121.6  & --   & 10.6 & \it {18.4} \\
1801$-$1417 &   3.625   &  --       &  57.2  & 31.5 & 17.5 & 23.7 \\
1802$-$2124 &  12.643   &   0.70    & 145.7  &  8.3 & 15.0 & \it {16.1} \\
1813$-$26   &   4.430   &  --       & 122.5  & 14.7 &  9.4 & 11.1 \\
1822$-$08   & 834.856   & 290.2     & 165.0  & 28.7 & 17.0 & 28.7 \\
1841+01     &  29.773   &  10.48    & 125.9  & 11.1 &  7.1 & 11.2 \\
1843$-$14   &   5.471   &  --       & 114.6  & 12.5 &  6.2 & 11.6 \\
1853+13     &   4.092   & 115.7     &  30.6  & 20.5 & 10.7 & 15.9 \\
1911+13     &   4.626   &  --       &  31.1  & 29.0 & 17.7 & 38.2 \\
1910+12     &   4.984   &  58.32    &  38.1  & 20.8 & 14.6 & 18.0 \\
\hline
\label{t:all_disc}
\end{tabular}
\end{center}
\small $^{\dag}$ Interim pulsar names have two digits of declination. \\
$^*$ The orbit for PSR J1723-28 has yet to be determined.
\end{small}
\end{table}

The detection of PSR B1635$-$45, (Johnston et al. 1992 \nocite{jlm+92})
was at a fundamental period of 529.1~ms instead of the catalogue period
of 264.55~ms. Further checking showed that the catalogue period was
incorrect and it has now been updated.

As noted in Section \ref{sec:processing}, filters are used to
eliminate some known interference signals. These filters can also
remove pulsars which happen to be at the same period. In the case of
PSR B1240$-$64, which has a spin frequency of 2.574Hz, its detection
was discarded, with the first 12 harmonics, by a filter which removed
frequencies from 0.6423Hz to 0.6443Hz together with the next 50
harmonics. Because PSR B1240$-$64 is very strong, having a S/N of
1466, it was seen at its 13th and other higher harmonics and strongly
at a period of 3/4 of the fundamental; however, a weaker pulsar might
have been lost.

Very strong pulsars are seen in multiple beams. One of the tests for
interference signals is that they are not localised to a small section
of sky, but seen in many of the beams on a tape. These signals
are excised from all the beams on a particular tape, assuming them to
be interference. This did not prove to be a problem for the PMPS,
since in all these cases the pulsar was seen in adjacent beams
on a different tape. However, at the periphery of a search the very
strongest pulsars could be lost.

\subsection{Multiple detections}

Reasonably strong pulsars or those near the edge of beams are likely
to have detections in adjacent beams. All these detections are
recorded for analysis and the strongest S/N listed as the `best
observation' as shown in Figure \ref{f:reap_flow}. \sc Reaper
\normalfont can relate beam position to pulsar position and any
anomalous behaviour can then be identified; for example, these
detections were used to improve the position of PSR J1744$-$2335 which
is close to the ecliptic plane, where timing alone does not give good
resolution in ecliptic latitude. In this case scintillation was not a
problem. Also, nulling pulsars can be identified, either by seeing
nulling within an observation or by finding that a pulsar is not
detected in the beam(s) in which a detection is expected.

A particular use of multiple detections is to extend the timing
baseline of newly discovered pulsars. After a year of timing
observations it is possible that a phase connected solution may be
found which will be coherent with the original detection
observation. Including a much earlier pulse arrival time will add
significant constraints to the ephemeris of the new pulsar. Ideally,
there will be multiple detections in the survey, which can be
identified using \sc Reaper \normalfont to further improve the timing
solution. We note that major surveys using multiple beam receivers use
interlaced observations to have complete coverage of an area. For
future timing of new discoveries using the survey observations, the
scheduling of adjacent beams to be evenly spread over the survey
period is worth considering.

\section{Discoveries}
\label{sec:discoveries}

\begin{figure*}
\begin{tabular}{cc}
\psfig{file=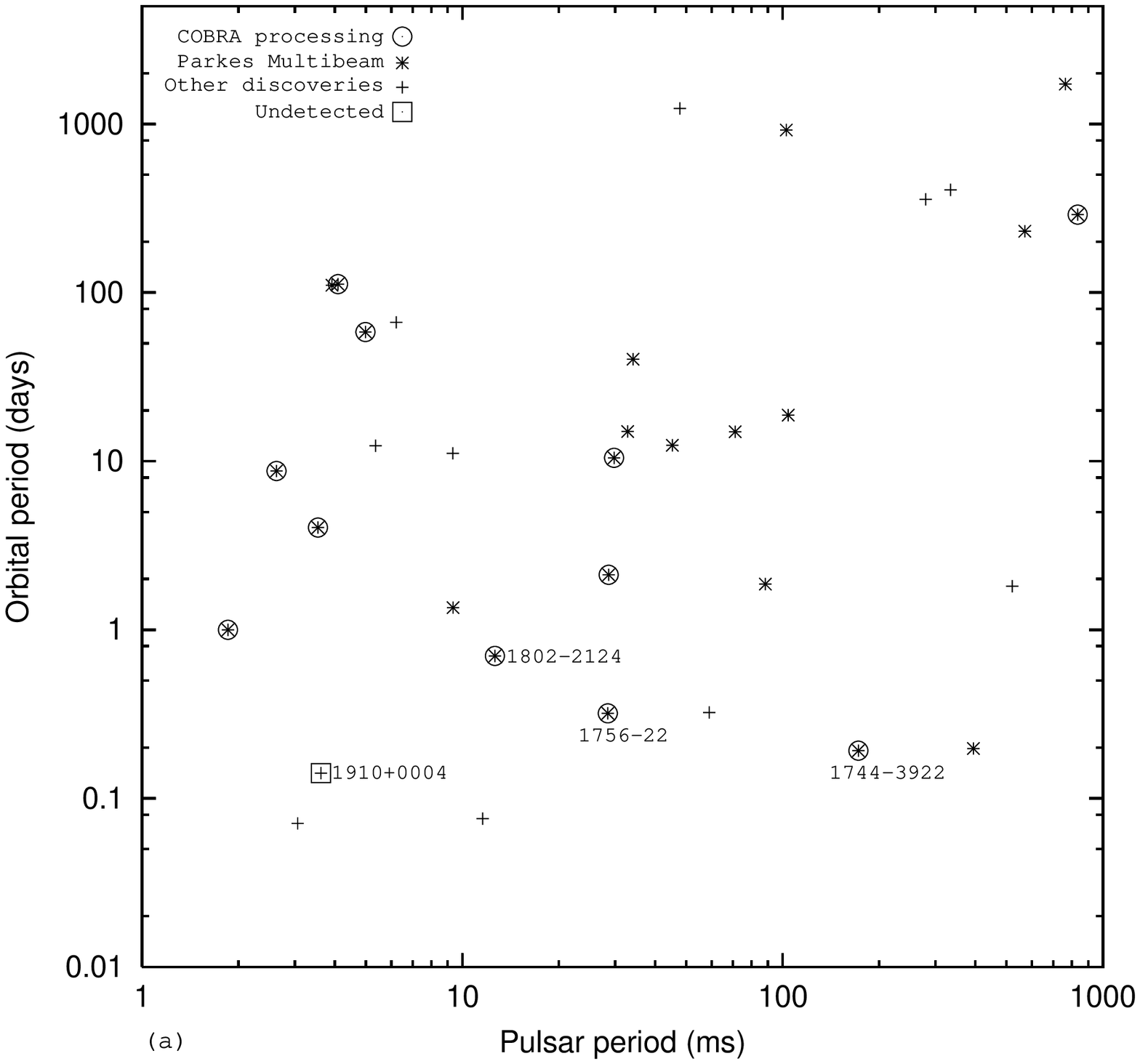,height=8.1truecm,angle=270} &
\psfig{file=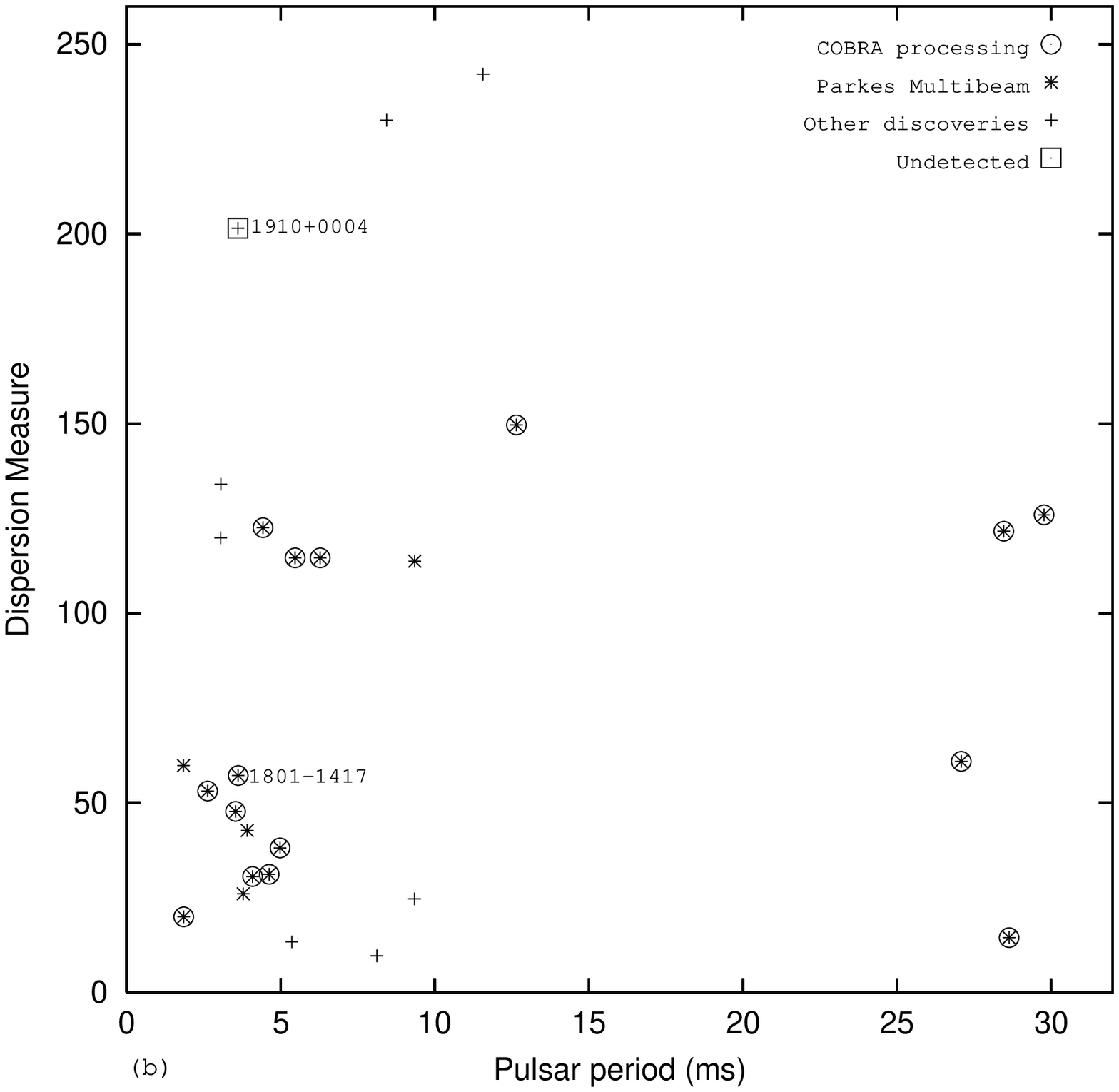,height=8.1truecm,angle=270}\\
\vspace{0.0truecm}
\end{tabular}
\caption{\small All binary and millisecond pulsars within the PMPS
search area. The systems discovered by the PMPS are shown with the new
discoveries enclosed in a circle. Panel (a) shows the relationship
between spin and orbital periods of the binary pulsars. Panel (b)
shows the spin period and DM for the millisecond pulsars. Named
pulsars are discussed in the text.}
\label{f:bin_milli}
\end{figure*}

There have been 128 new pulsars discovered in the COBRA processing.
These are being regularly observed to obtain a coherent timing
solution and they all will be published in future papers. A list of
provisional parameters for the eleven binary and fifteen millisecond
(spin period $<$ 30~ms) pulsar discoveries are shown in Table
\ref{t:all_disc}. All but two of the binary discoveries are also
millisecond pulsars.  Listed for comparison are the spectral S/N for
both the standard and stack acceleration searches and the S/N from the
candidate plot. For solitary and long orbital period binaries the S/N
for the stack search is lower than in the standard search, as
expected.  However, for the short spin period, tight orbit pulsars the
S/N with the stack search has been substantially improved; as has been
stated, for PSR J1756$-$22 there was no detection at all in the
standard search. It is worth noting that for very short spin periods
($<$ 10~ms), the spectral S/N is typically higher than the candidate
plot S/N which was derived in the time domain. This is due to the
small number of time samples over the period of the pulsar, making the
time domain S/N hard to determine accurately. \sc Reaper \normalfont
was used to investigate \itshape{spectral} \normalfont S/N against
period from $<$ 1~ms to 20~ms, specifically to look for short-period
candidates.

In Figure \ref{f:bin_milli} we plot spin period, orbital period and DM
of these new discoveries with the other known millisecond and binary
pulsars previously found in the search area. It can clearly be seen
that the COBRA processing, with improved interference filters and
acceleration searches, has been substantially more successful at
finding pulsars in binary systems with tight orbits and shorter pulsar
periods than the earlier PMPS processing. These objects are being
timed and will be discussed in future papers. Paper I discusses the
reduced sensitivity of the PMPS to pulsars with very short spin
periods at increasing DM just due to dispersion smearing. For example,
the sampling time is increased to 1.0~ms above a DM of
140~pc~cm$^{-3}$, to match the dispersion smearing and thus save
processing time. It can be seen in Figure \ref{f:bin_milli}(b) that
there are no PMPS millisecond pulsars above a DM of 150 pc~cm$^{-3}$.

\begin{figure}
\centerline{\psfig{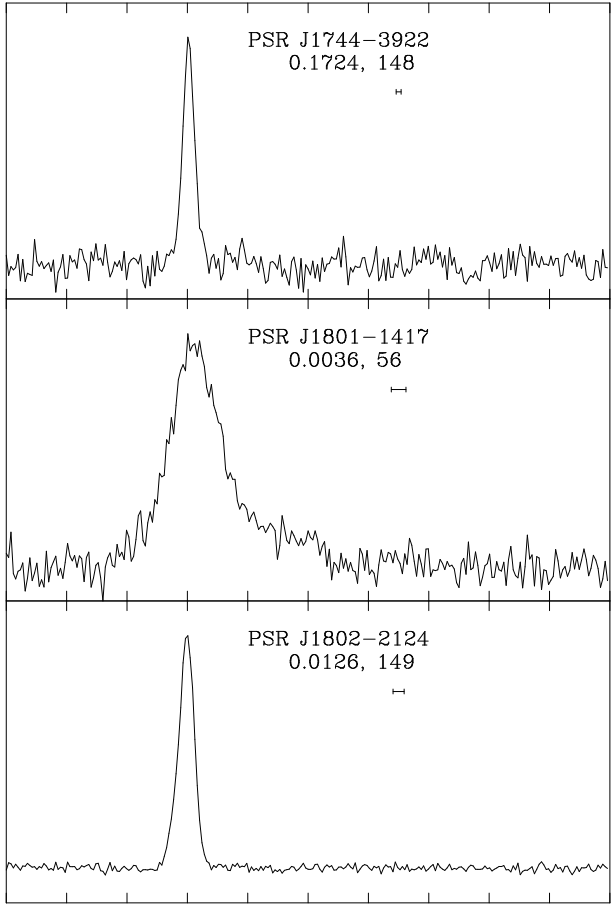}}
\vspace{0.0truecm}
\caption{\small Pulse profiles for the three pulsars discussed. The
highest point in the profile is placed at phase 0.3. The pulsar name,
period (in s) and DM (in pc~cm$^{-3}$) are given. The small horizontal
bar under the DM indicates the effective resolution of the
profile, including the effects of interstellar dispersion. Integration
times used were 313~min, 147~min and 599~min for PSRs J1744$-$3922,
J1801$-$1414 and J1802$-$2124, respectively.}
\label{f:profile_plot}
\end{figure}

In the following sub-sections we discuss three newly discovered
pulsars, two in binary systems and a solitary millisecond object. Both
of the binary systems have very low eccentricity; for such systems,
the longitude of periastron is not well defined. To avoid large
covariance between the usually quoted longitude and epoch of
periastron, we have adopted the ELL1 timing model, which uses the
Laplace--Lagrange parameters $\eta=\epsilon\sin\omega$ and
$\kappa=\epsilon\cos\omega$, where $\epsilon$ is the orbital
eccentricity and $\omega$ is the angle of periastron, plus the time of
ascending node T$_{\rm ASC}$. A detailed description of this binary
model can be found in Lange et al. (2001)\nocite{lcw+01}. The
parameters for the named pulsars are in Table \ref{t:complete},
including the derived parameters: characteristic age, $\tau_c=P/(2
\dot P)$; surface dipole magnetic field strength, $B_s=3.2\times
10^{19}(P\dot P)^{1/2}$~G. The pulsar mass was assumed to be
1.35~$M_{\odot}$. Distance, $d$, is computed from the DM with
both the TC93 model (Taylor \& Cordes 1993\nocite{tc93}), for
consistency with earlier publications, and NE2001 (Cordes \& Lazio
2002\nocite{cl02a}) for comparison. The pulse profiles are shown in
Figure \ref{f:profile_plot}. All of the pulsars reported here have had
regular timing observations covering more than one year, either at
Parkes or at Jodrell Bank Observatory or both. Timing analysis was
carried out using the \textsc{TEMPO} program\footnote{See
http://www.atnf.csiro.au/research/pulsar/tempo/} as described in Paper
I.

\begin{table*}
\begin{small}
\caption{\small Positions, flux densities, measured and derived
parameters for the newly discovered pulsars with complete timing
solutions. Values in parenthesis are twice the nominal \textsc{TEMPO}
uncertainties in the least significant digits quoted, obtained after
scaling TOA uncertainties to ensure $\chi^2_{\nu}=1$. The pulsar has
an assumed mass of 1.35$M_{\odot}$.}
\begin{center}
\begin{tabular}{llll}
\hline
\hline
Parameter	&
PSR J1744$-$3922	& PSR J1801$-$1417 & PSR J1802$-$2124 	\\
\hline
Right ascension (J2000) (h~~~m~~~s) &
17:44:02.675(10)& 18:01:51.0771(3) & 18:02:05.3352(3) 	\\
Declination (J2000) (~\degr ~~~\arcmin ~~~\arcsec)&
-39:22:21.1(4) 	& -14:17:34.547(3)  &-21:24:03.6(3)	\\

Galactic longitude (\degr) &
350.91		& 14.55		& 8.38		\\
Galactic latitude (\degr) &
-5.15		& +4.16		& +0.61		\\

Period $P$ (ms) 	& 
172.444360995(2)& 3.6250966597450(15) & 12.647593582763(5)  \\
Period derivative $\dot{P}$ (x $10^{-18}$) &
1.55(12)	& 0.00528(5)	& 0.072(1)	\\

Epoch (MJD)	& 
52530.0000	& 52340.0000	& 52855.0000	\\
Dispersion Measure DM (pc cm$^{-3}$) &
148.1(7)	& 57.2(1)	& 149.6(1)	\\

Orbital Period $P_b$(days) &
0.19140624(7) 	& ---		& 0.6988892434(8) \\
a sin i (lt-s) &
0.2120(2)	& ---		& 3.718866(7)	\\
\\
Laplace--Lagrange parameters: 			\\
$\eta$          &
0.8(20)$\times10^{-3}$ 	& ---	& 0.3(8) $\times10^{-5}$ \\
$\kappa$	&
0.1(2)$\times10^{-2}$ 	&  ---	& 0.1(6) $\times10^{-5}$ \\
T$_{\rm ASC}$ (MJD)	&
52927.16723(3) 	& ---		& 52595.7950781(4) \\
\\
Eccentricity $e$	&
$<$0.006	& ---		& $<$0.00001	\\

N$_{\rm toa}$	& 
36		& 64		& 32		\\
Timing data span (MJD)	&
51953 -- 53107 	& 51561 -- 53119 & 52605 -- 53105 \\
RMS timing residual ($\mu s$)	&
197		& 26.0		& 7.0		\\

Flux density at 1400MHz (mJy) &
0.20(3)		& 0.87(10)	& 0.77(9)	\\
Luminosity, log $Sd^2$ (mJy kpc$^2$) &
0.62		& 0.45		& 0.92  	\\
$W_{50}$ (ms)	&
3.4		& 0.35		& 0.37		\\
$W_{10}$ (ms)	&
---		& ---		& 0.74		\\

Characteristic Age $\tau_c$ (Gyr) &
1.8		& 10.8		& 2.8		\\
Magnetic Field $B_s$ ($10^8$ G) &
166		& 1.4		& 9.7		\\
Minimum companion mass ($M_{\odot}$)&
0.083         	& ---		& 0.81		\\

Distance $d$ (kpc) - TC93	&
4.6		& 1.8		& 3.3		\\
Distance $d$ (kpc) - NE2001	&
3.1		& 1.5		& 2.9		\\
$|z|$ (kpc)	&		
0.41 		& 0.13		& 0.04		\\
Detection beam number	&		
4		& 9		& 4		\\
Radial distance	(beam radii, $\sim$7~arcmin) &
1.3		& 0.8		& 0.9		\\

Detection signal-to-noise ratio (S/N) &
26.5		& 23.7		& 16.1		\\

\hline
\label{t:complete}
\end{tabular}
\end{center}
\end{small}
\end{table*}

\subsection{PSR J1744$-$3922}

PSR J1744$-$3922 was discovered in February 2003; it completes more
than 12\% of its orbit during a 35~min search
observation. Consequently, the acceleration correction improved the
optimised S/N by nearly 50\% from 17.8 to 26.5. There have been
regular observations at Parkes since the discovery. Subsequently,
pulse times of arrival from survey observations with detections of PSR
J1744$-$3922, taken in February and June 2001 have been used to
improve the timing solution substantially.  The companion appears to
be light, with a minimum mass of 0.083 solar masses. This pulsar has
an unusual combination of characteristics: (a) it is in a compact
circular orbit of only 4.6 hours with a projected semi-major axis of
0.21~lt-sec; (b) it has an unusually long spin period of 172~ms and;
(c) it experiences a lack of detectable emission for some of the
time. The spin period would suggest a young pulsar, however, the
characteristic age of 1.8 Gyr and a low eccentricity orbit indicates a
mildly recycled object.

The evolution of this type of system is the source of considerable
discussion in the literature; however a typical scenario for a low
mass binary pulsar with $P_b<1$ day is given by Phinney \&
Kulkarni (1994)\nocite{pk94}. In summary, the larger, primary star
with mass greater than the critical mass for a supernova explosion
($\sim8$ $M_{\odot}$) evolves and expands causing the secondary to
spiral-in. The He core of the primary eventually explodes forming a
neutron star. During the evolution of the secondary, mass transfer to
the neutron star takes place, thus spinning up the pulsar. The second
mass transfer phase circularises the orbit and probably causes a
second spiral-in phase, hence, the very tight orbit.

The position of PSR J1744$-$3922 is well within the 95\% error box of
EGRET source 3EG J1744$-$3934 (Hartman et al. 1999
\nocite{hbb+99}). Indeed, this pulsar was independently detected in a
targeted search of EGRET sources (Hessels et
al. 2004\nocite{hrs+04}). However, since the source is highly variable
(variability index, $V$, of 5.12, McLaughlin 2001 \nocite{mcl01}) and
sources due to pulsars typically have $V<1$, the association is
unlikely. Kramer et al. (2003) \nocite{kbm+03} reviewed possible PMPS
discovery associations with EGRET $\gamma$-ray sources. We follow the
same assessment approach using the pulsar's spin down energy given by:
$\dot{E}=4\pi^2I\dot PP^{-3}$ in erg s$^{-1}$, where a neutron-star
moment of inertia $I=10^{45}$ g cm$^{-2}$ is assumed. For PSR
J1744$-$3922, $\dot E=19.68\times10^{30}$erg s$^{-1}$ at a distance,
$d$, of 3.1kpc (using NE2001, Cordes \& Lazio 2002\nocite{cl02a}) such
that log[$\dot E/d^2$ (erg s$^{-1}$ kpc$^{-2}$)]=30.35. 3EG
J1744$-$3934 has a $\gamma$-ray flux, $\bar{F}$, of
$1.27\pm0.26\times10^{-10}$erg s$^{-1}$ cm$^{-2}$ implying a
$\gamma$-ray luminosity which is 550 times the spin-down
luminosity. Clearly, the EGRET source cannot be identified with the
spin-powered pulsar PSR J1744$-$3922.

The occasional lack of radio emission could be intrinsic or extrinsic
to the pulsar. Possible extrinsic phenomena are scintillation
(e.g. \nocite{lr68} Lyne \& Rickett 1968) or the companion star
obscuring the pulsar. There is no evidence of systematic changes in
the radiation characteristics at different phases of the orbit, so it
is unlikely to be due to eclipsing by the companion. With a DM of
148~pc~cm$^{-3}$ and an observing frequency of 1.4~GHz, we expect
strong diffractive scintillation to produce scintles with a typical
width of a few MHz (Cordes et al. 1985\nocite{cwb85}). The bandwidth
of the receiving system is 288~MHz (see Paper I), consequently, any
strong scintillation effects are averaged out and would not be
seen. It is likely that the cause of the radiation fluctuations is
intrinsic and probably due to nulling, first reported by Backer
(1970)\nocite{bac70} for normal pulsars.


The nulling of PSR J1744-3922 occurs in short timescales. The pulse
can be visible from less than a minute to a few minutes with off-times
up to a few tens of minutes. Some 35~min observations do not have any
detections. Overall, there is no detectable radiation at 20~cm for
$\sim$75\% of the time. To our knowledge this is the first time that
nulling has been observed in a binary or a (mildly) recycled
pulsar.

There have been attempts to find correlations between pulsar nulling
and other pulsar characteristics (e.g. Rankin 1986 \nocite{ran86} and
Biggs 1992\nocite{big92}); for instance, Biggs found a marginally
significant correlation between pulsar characteristic age and null
fraction. For recycled pulsars many parameters are considerably
different e.g. they have much larger characteristic ages and their
periods are typically shorter. Until this discovery, recycled pulsars
were not observed to null. Physically, it is possible that the
potential drop, $\Delta\Psi$, between the magnetic pole and polar cap
may have an influence on a pulsar's emission mechanism. We have
investigated the magnitude of this potential for different types of
pulsars to look for evidence of a correlation. In terms of observable
parameters, the potential drop:
\begin{equation}
\Delta\Psi\simeq2\times 10^{13}
\left(\frac{P}{s}\right)^{-3/2} 
\left(\frac{\dot P}{10^{-15}}\right)^{1/2}{\rm~V},
\end{equation}
where $P$ is spin period and $\dot P$ the period derivative (Goldreich
\& Julian 1969\nocite{gj69}). We considered the median potential drops
for the groups of pulsars and compared the results with PSR
J1744$-$3922. The results are shown in Table \ref{t:null}. It can be
seen that nulling pulsars may have a lower potential drop than the
median of all pulsars or all recycled pulsars, which we have somewhat
arbitrarily defined as having a characteristic age $>10^7$~yrs and
surface magnetic field $<10^{11}$~G. There is a difference between
recycled pulsars with $P<100$~ms and those with $P>100$~ms, which we
show separately in Table \ref{t:null}. The lowest potential drop is
for recycled pulsars with $P\geq100$~ms, which is the category for PSR
J1744$-$3922. With some exceptions, there could be a trend among
normal pulsars between nulling and $\Delta\Psi$; the same relationship
between PSR J1744$-$3922 and other similar objects is not clear.

\vspace{0.5truecm}

\nocite{mht+04} \nocite{ran86} \nocite{big92}

\begin{table}
\begin{small}
\caption{\small The potential drop $\Delta\Psi$ for groups of
pulsars. All the published data are taken from the ATNF Catalogue
(Manchester et al. 2004). The groups used are: [1] all pulsars; [2]
pulsars known to null taken from Rankin (1986), Biggs (1992) and
PMPS pulsars observed to null; [3] all recycled pulsars (magnetic
field $<10^{11}$~G and characteristic age $>10^8$~yrs); [4]
short-period recycled pulsars ($P<100$~ms); long-period recycled
pulsars ($P\geq100$~ms). Included is PSR J1744$-$3922 for comparison.}

\begin{center}
\begin{tabular}{lrlr}
\hline
\hline
Type of Pulsar	& Number    	& \multicolumn{1}{c}{log[$\Delta\Psi$(V)]} & \\
                & 	 	& Median	& \\
\hline
\vspace{0.2cm}
All pulsars	& 1381		& $13.8^{+0.7}_{-0.7}$ 	& [1]\\
\vspace{0.1cm}
Known nullers	& 29  		& $13.2^{+0.4}_{-0.1}$ 	& [2]\\
\vspace{0.1cm}
All recycled	& 121 		& $14.1^{+0.6}_{-1.0}$ 	& [3]\\
\vspace{0.1cm}
`short' recycled & 79  		& $14.3^{+0.6}_{-0.4}$ 	& [4]\\
\vspace{0.1cm}
`long' recycled	 & 42  		& $13.0^{+0.3}_{-0.3}$  & [5]\\
\vspace{0.1cm}
PSR J1744$-$3922 & 1 		& $13.0$		&    \\
\hline
\label{t:null}
\end{tabular}
\end{center}
\end{small}
\end{table}

\subsection{PSR J1802$-$2124}

This system has a heavy companion with a minimum mass $>$0.8
$M_{\odot}$ which could be another neutron star. However, the
circular orbit of the pulsar implies that it is unlikely that a second
supernova explosion occurred in the system, hence the companion is
probably a heavy white dwarf (e.g. Phinney \& Kulkarni
1994\nocite{pk94}). The evolution is likely to be similar to that of
PSR J1744$-$3922, except that the progenitor secondary star was
considerably more massive in this system. This would result in the
neutron star being engulfed during the red giant phase of the
companion, accreting material and consequently being spun-up. This
process would also cause the two stars to spiral-in towards each other
and to eject the giant's envelope, leaving a pulsar in a tight, circular
orbit with a large white dwarf companion. The system is categorised as
being an intermediate mass binary pulsar (IMBP) as introduced by
Camilo et al. (1996)\nocite{cnst96} and further discussed in Camilo et
al. (2003)\nocite{clm+01}, defined as having intermediate
mass donor stars. They identified binary systems with the pulsar spin
period 10$< P <$200 ms, $e < 0.001$, companion mass $>$0.4M$_{\odot}$
and 0.5$<$P$_b<$15~days as likely IMBP's.  PSR J1802$-$2124 meets all
these criteria and is clearly a member of this class of system; it has
added to the sub-class of relatively short spin periods with a 12.6~ms
rotation, similar to PSR's J1435$-$6100/J1757$-$5322 which contrast
with PSR B0655+54 at 195~ms noted by Camilo et al.

PSR J1802$-$2124 brings the number of IMBPs known to 15 (for a list of
the other 14 see Hobbs et al. 2004\nocite{hfs+04}); with eight either
found or detected by the PMPS. This concentration near to the Galactic
plane is discussed in Camilo et al. (2001), with a possible reason
being the birth velocity is restricted because of the high system
mass. PSR J1802$-$2124 with $|z|$ of only 0.04~kpc supports this view,
although there may be selection effects due to the high sensitivity
survey coverage of the PMPS along the Galactic plane.

PSR J1802$-$2124 has a narrow pulse profile of $<$ 0.4~ms ($<$ 3\% of
the period), which may be even narrower if the effects of dispersion
smearing are removed. This probably accounts for the flux density
being apparently less than the nominal PMPS flux limit, which is
calculated at a 5\% duty cycle.

\subsection{PSR J1801$-$1417 }

PSR J1801$-$1417 is a solitary 3.62~ms pulsar first discovered in
November 2002. It has a characteristic age of 10.8 Gyr. The ephemeris
of this pulsar has been considerably improved by including three
detections from the PMPS back to January 2000.  

Following an earlier suggestion by Bailes et
al.~(1997)\nocite{bjb+97}, Kramer et al. (1998)\nocite{kxc+98} put
forward evidence that solitary millisecond pulsars are less luminous
than when they are in binary systems. By calculating the flux density
times the square of the distance, $Sd^2$, we can make luminosity
comparisons; the values of log $Sd^2$ (mJy kpc$^2$) for our sources
are shown in Table \ref{t:complete}. As can be seen the luminosity of
PSR J1801$-$1417 is significantly lower than that of PSR J1802$-$2124,
although both are more luminous than the ranges identified by Kramer
et al. (mean value, log $Sd^2$ of $-0.5\pm0.3$ and $0.2\pm0.1$ for
solitary and binary millisecond pulsars respectively). We note that it
is unlikely that either of these pulsars would have been discovered if
their luminosity was as low as Kramer et al.'s ranges, both pulsars
exceed the 1.5~kpc previously used, with PSR J1802$-$2124 at 3.3~kpc.

We have updated the luminosity statistics for recycled pulsars, as
defined by Kramer et al. (1998) for the current known population using
a distance limit of 2.0~kpc, increased from 1.5~kpc. The population
has increased from 7 to 10 solitary pulsars and from 11 to 19 in
binary systems. With a distance limit of 2.0~kpc the luminosity mean
values are log $Sd^2$ of $-0.4\pm0.2$ and $0.1\pm0.1$ for solitary and
binary objects respectively.  These results, which were also checked
at a limit of 1.5~kpc, are consistent with Kramer et al. and provide
further evidence that solitary millisecond pulsars are less luminous
radio sources than their binary counterparts.  The cause of this
effect is still not known. There will be future analysis of all the
new millisecond discoveries of the PMPS.

\section{Future Work}
\label{sec:discussion}

The PMPS has been the most successful search for pulsars ever
undertaken. However, there is still further work to do. The processing
performed so far is still insensitive to fast millisecond pulsars in
binary orbits of a few hours or less; for example the binary pulsar
PSR J0737$-$3039 (Burgay et al. 2003\nocite{bdp+03}), would only be
detected at the times in its 2.4~hour orbit where the apparent spin
period changes are most constant. This system was found using
essentially the same software system as used in this processing, but
the observation time was only four minutes, which radically reduces
pulse smearing. Short observations also reduce potential sensitivity,
so, there is considerable scope for further reanalysis of the data
with more sophisticated software or as partial observations.

We are processing the observations in four quarters, using the
standard and stack search algorithms. With a shorter observation time
there will be less pulse smearing of highly accelerated pulsars as
discussed in Section \ref{s:acceleration}. This may find objects in
fast orbits, but at a price of reduced base sensitivity by a factor of
2. Evolutionary population studies suggest that black hole - pulsar
binaries exist in the Galactic field (Sigurdsson 2003\nocite{sig03}).
Such systems may be in very tight orbits. One or more could be in the
PMPS dataset and a better acceleration search algorithm may find them.

\section{Acknowledgements} 

 We gratefully acknowledge the technical assistance with hardware and
 software provided by Jodrell Bank Observatory, CSIRO ATNF,
 Osservatorio Astronomico di Bologna, Swinburne centre for
 Astrophysics and Supercomputing.  The Parkes radio telescope is part
 of the Australia Telescope which is funded by the Commonwealth of
 Australia for operation as a National Facility managed by CSIRO. IHS
 holds an NSERC UFA and is supported by a Discovery Grant. DRL is a
 University Research Fellow funded by the Royal Society.  FC
 acknowledges support from NSF grant AST-02-05853 and a NRAO travel
 grant. NDA, AP and MB received support from the Italian Ministry of
 University and Research (MIUR) under the national program {\it Cofin
 2002}.

\bibliographystyle{mn}

\begin{thebibliography}{{{Ransom}, {Eikenberry} \& {Middleditch} }{2002}}

\bibitem[\protect\citename{Andersson, Kokkotas \& Schutz }{1999}]{aks99}
Andersson~N., Kokkotas~K., Schutz~B.~F., 1999, ApJ, 510, 846

\bibitem[\protect\citename{Backer }{1970}]{bac70}
Backer~D.~C., 1970, Nature, 228, 42

\bibitem[\protect\citename{Backer {\rm et~al. }}{1982}]{bkh+82}
Backer~D.~C., Kulkarni~S.~R., Heiles~C., Davis~M.~M., Goss~W.~M., 1982, Nature,
  300, 615

\bibitem[\protect\citename{Bailes {\rm et~al. }}{1997}]{bjb+97}
Bailes~M. {\rm et~al.}, 1997, ApJ, 481, 386

\bibitem[\protect\citename{Biggs }{1992}]{big92}
Biggs~J.~D., 1992, in Hankins~T.~H., Rankin~J.~M., Gil~J.~A., eds, The
  Magnetospheric Structure and Emission Mechanisms of Radio Pulsars, {IAU}
  Colloquium 128.
\newblock Pedagogical University Press, Zielona G\'ora, Poland, p.~22

\bibitem[\protect\citename{Burgay }{2004}]{bur04}
Burgay~M., 2004, {\rm PhD thesis}, The University of Bologna

\bibitem[\protect\citename{{Burgay} {\rm et~al. }}{2003}]{bdp+03}
{Burgay}~M. {\rm et~al.}, 2003, Nature, 426, 531

\bibitem[\protect\citename{Camilo {\rm et~al. }}{1996}]{cnst96}
Camilo~F., Nice~D.~J., Shrauner~J.~A., Taylor~J.~H., 1996, ApJ, 469, 819

\bibitem[\protect\citename{{Camilo} {\rm et~al. }}{2000a}]{clf+00}
{Camilo}~F., {Lorimer}~D., {Freire}~P., {Lyne}~A.~G., {Manchester}~R.~N.,
  2000a, ApJ, 535, 975

\bibitem[\protect\citename{Camilo {\rm et~al. }}{2000b}]{clm+00}
Camilo~F. {\rm et~al.}, 2000b, in Kramer~M., Wex~N., Wielebinski~R., eds,
  Pulsar Astronomy - 2000 and Beyond, {IAU} Colloquium 177.
\newblock Astronomical Society of the Pacific, San Francisco, p.~3,
  astro-ph/9911185

\bibitem[\protect\citename{Camilo {\rm et~al. }}{2001}]{clm+01}
Camilo~F. {\rm et~al.}, 2001, ApJ, 548, L187

\bibitem[\protect\citename{{Cordes} \& {Lazio} }{2002}]{cl02a}
{Cordes}~J.~M., {Lazio}~T.~J.~W., 2002, {astro-ph/0207156}

\bibitem[\protect\citename{Cordes, Weisberg \& Boriakoff }{1985}]{cwb85}
Cordes~J.~M., Weisberg~J.~M., Boriakoff~V., 1985, ApJ, 288, 221

\bibitem[\protect\citename{Damour \& {Esposito-Far{\` e}se} }{1998}]{de98}
Damour~T., {Esposito-Far{\` e}se}~G., 1998, Phys. Rev. D, 58(042001), 1

\bibitem[\protect\citename{Dhurandhar \& Vecchio }{2001}]{dv01}
Dhurandhar~S.~V., Vecchio~A., 2001, Phys. Rev. D, 63(12), 122001

\bibitem[\protect\citename{{Edwards} {\rm et~al. }}{2001}]{ebsb01}
{Edwards}~R.~T., {Bailes}~M., {van Straten}~W., {Britton}~M.~C., 2001, MNRAS,
  326, 358

\bibitem[\protect\citename{Friedman }{1995}]{fri95}
Friedman~J.~L., 1995, in Fruchter~A.~S., Tavani~M., Backer~D.~C., eds,
  Millisecond Pulsars: A Decade of Surprise.
\newblock Astronomical Society of the Pacific Conference Series, p.~177

\bibitem[\protect\citename{Goldreich \& Julian }{1969}]{gj69}
Goldreich~P., Julian~W.~H., 1969, ApJ, 157, 869

\bibitem[\protect\citename{Hartman {\rm et~al. }}{1999}]{hbb+99}
Hartman~R.~C. {\rm et~al.}, 1999, ApJS, 123, 79

\bibitem[\protect\citename{{Hessels} {\rm et~al. }}{2004}]{hrs+04}
{Hessels}~J.~W.~T., {Ransom}~S.~M., {Stairs}~I.~H., {Kaspi}~V.~M.,
  {Freire}~P.~C.~C., {Backer}~D.~C., {Lorimer}~D.~R., 2004, in IAU Symposium.
\newblock p.~131

\bibitem[\protect\citename{Hobbs }{2002}]{hob02}
Hobbs~G., 2002, {\rm PhD thesis}, University of Manchester

\bibitem[\protect\citename{{Hobbs} {\rm et~al. }}{2004}]{hfs+04}
{Hobbs}~G. {\rm et~al.}, 2004, MNRAS, , in press

\bibitem[\protect\citename{Johnston \& Kulkarni }{1991}]{jk91}
Johnston~H.~M., Kulkarni~S.~R., 1991, ApJ, 368, 504

\bibitem[\protect\citename{Johnston {\rm et~al. }}{1992}]{jlm+92}
Johnston~S., Lyne~A.~G., Manchester~R.~N., Kniffen~D.~A., D'Amico~N., Lim~J.,
  Ashworth~M., 1992, MNRAS, 255, 401

\bibitem[\protect\citename{{Jouteux} {\rm et~al. }}{2002}]{jrs+02}
{Jouteux}~S., {Ramachandran}~R., {Stappers}~B.~W., {Jonker}~P.~G., {van der
  Klis}~M., 2002, A\&A, 384, 532

\bibitem[\protect\citename{Kramer {\rm et~al. }}{1998}]{kxc+98}
Kramer~M., Xilouris~K.~M., Lorimer~D., Doroshenko~O., Jessner~A.,
  Wielebinski~R., Wolszczan~A., Camilo~F., 1998, ApJ, 501, 270

\bibitem[\protect\citename{{Kramer} {\rm et~al. }}{2003}]{kbm+03}
{Kramer}~M. {\rm et~al.}, 2003, MNRAS, 342, 1299

\bibitem[\protect\citename{Lange {\rm et~al. }}{2001}]{lcw+01}
Lange~C., Camilo~F., Wex~N., Kramer~M., Backer~D., Lyne~A., Doroshenko~O.,
  2001, MNRAS, 326, 274

\bibitem[\protect\citename{Lyne \& Rickett }{1968}]{lr68}
Lyne~A.~G., Rickett~B.~J., 1968, Nature, 218, 326

\bibitem[\protect\citename{Manchester {\rm et~al. }}{2001}]{mlc+01}
Manchester~R.~N. {\rm et~al.}, 2001, MNRAS, 328, 17

\bibitem[\protect\citename{{Manchester} {\rm et~al. }}{2004}]{mht+04}
{Manchester}~R., {Hobbs}~G., {Teoh}~A., {Hobbs}~M., 2004, ApJ, submitted

\bibitem[\protect\citename{{McLaughlin} \& {Cordes} }{2003}]{mc03}
{McLaughlin}~M.~A., {Cordes}~J.~M., 2003, ApJ, 596, 982

\bibitem[\protect\citename{McLaughlin }{2001}]{mcl01}
McLaughlin~M., 2001, {\rm PhD thesis}, Cornell University

\bibitem[\protect\citename{{Morris} {\rm et~al. }}{2002}]{mhl+02}
{Morris}~D.~J. {\rm et~al.}, 2002, MNRAS, 335, 275

\bibitem[\protect\citename{Phinney \& Kulkarni }{1994}]{pk94}
Phinney~E.~S., Kulkarni~S.~R., 1994, Ann. Rev. Astr. Ap., 32, 591

\bibitem[\protect\citename{Rankin }{1986}]{ran86}
Rankin~J.~M., 1986, ApJ, 301, 901

\bibitem[\protect\citename{{Ransom}, {Cordes} \& {Eikenberry} }{2003}]{rce03}
{Ransom}~S.~M., {Cordes}~J.~M., {Eikenberry}~S.~S., 2003, ApJ, 589, 911

\bibitem[\protect\citename{{Ransom}, {Eikenberry} \& {Middleditch}
  }{2002}]{rem02}
{Ransom}~S.~M., {Eikenberry}~S.~S., {Middleditch}~J., 2002, AJ, 124, 1788

\bibitem[\protect\citename{{Sigurdsson} }{2003}]{sig03}
{Sigurdsson}~S., 2003, in Bailes~M., Nice~D.~J., Thorsett~S., eds, ASP Conf.
  Ser. 302: Radio Pulsars.
\newblock p.~391

\bibitem[\protect\citename{Staelin }{1969}]{sta69}
Staelin~D.~H., 1969, Proc. I. E. E. E., 57, 724

\bibitem[\protect\citename{Taylor \& Cordes }{1993}]{tc93}
Taylor~J.~H., Cordes~J.~M., 1993, ApJ, 411, 674

\bibitem[\protect\citename{Taylor \& Huguenin }{1969}]{th69}
Taylor~J.~H., Huguenin~G.~R., 1969, Nature, 221, 816

\bibitem[\protect\citename{Taylor \& Weisberg }{1989}]{tw89}
Taylor~J.~H., Weisberg~J.~M., 1989, ApJ, 345, 434

\bibitem[\protect\citename{Taylor }{1993}]{tay93a}
Taylor~J.~H., 1993, in Fontaine~G., V\^an~J. T.~T., eds, Particle Astrophysics,
  {IV}th {R}encontres de {B}lois.
\newblock Editions Frontieres, Gif-sur-Yvette, France, p.~367

\bibitem[\protect\citename{{Wood} {\rm et~al. }}{1991}]{wnh+91}
{Wood}~K.~S. {\rm et~al.}, 1991, ApJ, 379, 295

\end{thebibliography}

\end{document}